\documentclass[paper,prd,twocolumn,nofootinbib,preprintnumbers,amsmath,amssymb]{revtex4-2}

\usepackage{latexsym,graphicx,amssymb,amsmath,mathrsfs,bbold} 
\usepackage{slashed,mathbbol}
\usepackage[utf8]{inputenc}

\newcommand{\be}{\begin{equation}}      
\newcommand{\ee}{\end{equation}}      
\newcommand{\bea}{\begin{eqnarray}}      
\newcommand{\eea}{\end{eqnarray}}    
\newcommand{\Tr}{\,\textrm{Tr}\,}
\newcommand{\un}{\mbox{$\mathbb{1}$}}

\makeatletter
\renewcommand\appendix{\par
\setcounter{section}{0}%   
\setcounter{subsection}{0}% 
\gdef\thesection{\appendixname\space\@Alph\c@section}}

\long\def\unmarkedfootnote#1{{\long\def\@makefntext##1{##1}\footnotetext{#1}}}
\makeatother

\begin{document} 

\title{Second-order chiral phase transition in three-flavor quantum chromodynamics?} 
\author{G. Fej\H{o}s}
\email{gergely.fejos@ttk.elte.hu}
\affiliation{Institute of Physics, E\"otv\"os University, 1117 Budapest, Hungary}

\begin{abstract}
{We calculate the renormalization group flows of all perturbatively renormalizable interactions in the three-dimensional Ginzburg-Landau potential for the chiral phase transition of three-flavor quantum chromodynamics. On the contrary to the common belief we find a fixed point in the system that is able to describe a second-order phase transition in the infrared. This shows that long-standing assumptions on the transition order might be false. If the transition is indeed of second-order, our results may hint that the axial $U(1)$ symmetry restores at the transition temperature.}
\end{abstract}

\maketitle

\section{Introduction}

The nature of the chiral phase transition of quantum chromodynamics (QCD) with massless quarks is a  question that has been under debate for decades. The seminal work of Pisarski and Wilczek \cite{pisarski84} showed that when applying the $\epsilon$ expansion to the renormalization group (RG) flows of the Ginzburg-Landau potential of the chiral transition, no infrared (IR) stable fixed point appears, indicating that irrespectively of the number of quark flavors, $N_f$, the transition is of first-order. This analysis was based on the assumption that at the transition point the axial anomaly disappears, which, if, in fact does survive the thermal evolution, then according to \cite{pisarski84}, for $N_f=2$, the transition becomes of second-order with $O(4)$ critical exponents. For $N_f\geq 3$, however, the anomaly does not appear to have significance and either way one arrives at the conclusion of a first-order transition, even after including higher loop effects \cite{butti03}.

Studies using the functional renormalization group (FRG) also seemed to have confirmed the results of the $\epsilon$ expansion. Numerous papers reported fluctuation-induced first-order transitions for various flavor numbers; see $N_f=2$ \cite{fukushima11,grahl13}, $N_f=3$ \cite{resch19}, $N_f\geq 2$ \cite{fejos14}. These studies typically neglected the momentum dependence of the $n$-point vertices, i.e., used a local potential approximation, which was reasonable for evaluating the effective action in homogeneous backgrounds (thus investigating the transition order), but they generally approximated the local potential via a restricted set of chirally symmetric operators. Still, apart from the case of a substantial axial anomaly for $N_f=2$, all analyses agree that the only way to obtain a second-order transition is to entirely drop fluctuations, i.e., employ Landau's theory. If explicit fermion degrees of freedom are also introduced into the effective description, then in the mean-field approximation (i.e., the sole inclusion of one-loop fermion terms), the second-order prediction of Landau's theory survives \cite{resch19}. The results of \cite{pisarski84}, however, show that fluctuations of the chiral field must not be dropped, but then one inevitably seems to get first-order transitions.

It has to be noted that for $N_f=2$, FRG explicitly confirmed that an $O(4)$ fixed point can indeed be hit in the IR, but only if the anomaly is strong enough \cite{grahl13}. Furthermore, (for $N_f=2$) there have been indications that the transition could be of second-order even if the axial $U(1)$ symmetry does get restored at the transition temperature \cite{pelissetto13,grahl14}, but then presumably belonging to a different universality class. That is, for $N_f=2$, fluctuations of the chiral order parameter may form a new IR fixed point, but no convincing evidence emerged that it could indeed be physical. No similar results were obtained for $N_f\geq 3$ and it has even become textbook material \cite{yagibook} that the transition for $N_f=3$ is discontinuous in the chiral limit, which may or may not extend to $N_f=2$, depending on the thermal fate of the axial anomaly.

Earlier attempts using lattice QCD simulations also agreed with the first-order nature of the chiral transition \cite{karsch01,deforcrand03,jin15}, but cutoff effects were typically large and it was also argued that the existence of a discontinuous transition might be questionable \cite{bazazov17}. The problem with lattice QCD is that due to the singular behavior of the fermion determinant in the zero quark mass limit, a direct approach for simulations is not available.  Furthermore, it is a notoriously hard task to acquire the chiral limit via a sequence of finite quark masses while keeping the cutoff and infinite volume limits under control. 

Thermal fate of the axial anomaly in the chiral limit is not settled either. There are numerous lattice studies from the past decade that support \cite{brandt16,tomiya16,aoki21}, and many that are against \cite{bazazov12,buchoff14,bhattacharya14,dick15,ding21,kaczmarek21} the restoration of the axial $U(1)$ symmetry at the transition point. In this study we attempt to argue that the transition order for $N_f=3$ could be very sensitive to the anomaly evolution toward the chiral transition, and exploration in this direction is of huge importance \cite{lahiri21}.

Even though most studies pointed in the direction of a discontinuous chiral transition for three flavors, recently, via lattice QCD simulations, Cuteri, Philipsen, and Sciarra conjectured that, for vanishing quark masses, the transition might become second-order \cite{cuteri21}. This claim was followed by \cite{dini22} obtaining a similar result. These lattice QCD studies are in serious conflict with all the renormalization group arguments, since for $N_f=3$ neither in the $\epsilon$ expansion, nor via the FRG the existence of an IR stable fixed point emerges, which is a necessary condition for critical behavior and a second-order transition. The main motivation of this study is to show via the FRG that this conflict may potentially be resolved.

\section{Ginzburg-Landau theory and \\renormalization group flows}

According to the Ginzburg-Landau paradigm, for physical systems close to a second-order transition, there exists a local order parameter, $\Phi$, emerging from some averaging of microscopic degrees of freedom, which can be used as an expansion parameter in the coarse-grained free energy, ${\cal F}$, at a suitable ultraviolet (UV) momentum scale. For the three-flavor chiral transition, $\Phi$ is a $3\times 3$ complex matrix, parametrized as $\Phi = \phi_a T_a \equiv (\sigma_a+i\pi_a)T_a$, where $T_a$ are the usual $U(3)$ generators,  $\Tr (T_aT_b)=\delta_{ab}/2$. One needs to write down in growing powers of $\phi_a$ the most general free-energy functional that respects $U(3)\times U(3)$ chiral symmetry, which acts as $\Phi \longrightarrow L \Phi R^\dagger$, where $L$ and $R$ are arbitrary $U(3)$ matrices. A possible set of independent combinations that can appear in ${\cal F}$ is
\bea
\label{Eq:Is}
I_1=\Tr \big(\Phi^\dagger \Phi\big), \quad I_2 = \Tr \big(\Phi^\dagger \Phi - \Tr(\Phi^\dagger \Phi)/3\cdot \un\big)^2, \nonumber\\
\hspace{-2cm}I_3 = \Tr \big(\Phi^\dagger \Phi - \Tr(\Phi^\dagger \Phi)/3\cdot \un\big)^3. \hspace{1.3cm}
\eea
It can be shown that any other chirally invariant term can be expressed as a function of $I_1$, $I_2$, and $I_3$. The axial anomaly is described by the Kobayashi--Maskawa--'t Hooft determinant,
\bea
\label{Eq:Idet}
I_{\det} = \det \Phi + \det \Phi^\dagger,
\eea
as it is invariant under any chiral transformation, except for the axial $U(1)$ subgroup. Note that $\tilde{I}_{\det} = \det \Phi - \det \Phi^\dagger$ is forbidden due to parity reasons, while $\tilde{I}_{\det}^2$ (and thus $\det \Phi^\dagger\cdot \det \Phi$) is not independent, as it can be expressed in terms of (\ref{Eq:Is}) and (\ref{Eq:Idet}). Therefore, the most general functional for ${\cal F}$ that respects chiral symmetry but breaks axial $U(1)$ is
\bea
\label{Eq:F}
{\cal F}[\Phi] &=& \int d^3x \Big [m^2 I_1 + aI_{\det}+ g_1 I_1^2 + g_2 I_2 +  bI_1 I_{\det}\nonumber\\
&&\hspace{0.9cm}+\lambda_1 I_1^3 + \lambda_2 I_1 I_2 + a_2 I_{\det}^2+g_3 I_3 + {\cal O}(\phi^7) \nonumber\\
&&\hspace{0.9cm}+\Tr[\partial_i \Phi^\dagger \partial_i \Phi] +{\cal O}(\partial^4 \phi^2)\Big].
\eea
Note that in (\ref{Eq:F}) we substantially extended the operator set compared to earlier treatments, and included all perturbatively renormalizable operators in three dimensions. This is in contrast with four dimensions (and thus with the $\epsilon$ expansion), where only the couplings $m^2$, $a$, $g_1$, and $g_2$ would appear. On top of these constants, we now have five more, i.e., $b$, $\lambda_1$, $\lambda_2$, $a_2$, $g_3$. 

To determine the scale ($k$) evolution of the free energy, ${\cal F}_k$, we use the FRG technique. This is particularly convenient as it allows the evaluation of all $\beta$ functions directly in three dimensions. The scale dependence is described by Wetterich's flow equation \cite{wetterich93,morris94}. Under the assumption that the momentum structure of ${\cal F}$ retains its UV form throughout the scale evolution (i.e., generation of non-renormalizable higher derivative terms are dropped), Litim's optimization procedure \cite{litim01} leads to the following form of the flow equation:
\bea
\label{Eq:flow}
k\partial_k {\cal F}_k[\Phi]&=& k^2 \int_{|\vec{q}|<k} \frac{d^3q}{(2\pi)^3 }\Tr (k^2\cdot \un + {\cal F}_k'')_{q=0}^{-1}[\Phi]\nonumber\\
&=&\frac{k^5}{6\pi^2}\Tr (k^2\cdot \un + {\cal F}_k'')_{q=0}^{-1}[\Phi],
\eea
where $\Phi$ is a homogeneous but otherwise arbitrary field configuration. As in ordinary Wilsonian renormalization, it is assumed that the scale evolution is governed by the $k$ dependence of the renormalizable couplings, i.e., effects of (perturbatively) irrelevant operators are dropped.  Note that these terms are still generated during the scale evolution and it is not known {\it a priori} whether they can alter the fixed-point structure and their stability. Further investigations on this issue, including the question of the wave-function renormalization and higher derivative terms are very important, but beyond the scope of this paper.

Flows of the couplings are obtained by $(i)$ calculating the zero momentum part of the $18\times 18$ second derivative matrix, ${\cal F}''_{k,q=0}$, in a homogeneous background $\Phi$, then $(ii)$ expanding the rhs of (\ref{Eq:flow}) in terms of the components of $\Phi$ leading to the emergence of all chiral invariants, and finally $(iii)$ matching all terms in the lhs and rhs of (\ref{Eq:flow}), which yields the $k$ derivative of each individual coupling. This leads to the $\beta$ functions, which, as usual, are defined as the logarithmic $k$ derivative of the dimensionless couplings, which are rescaled versions of the original couplings with appropriate powers of $k$. These dimensionless couplings will be denoted by a bar on top (e.g., $m_k^2=k^2\bar{m}_k^2$).

\section{$\beta$ functions and fixed points} 

Evaluating the rhs of (\ref{Eq:flow}) in a general background of $\Phi$ is problematic in its full generality, as it is practically impossible to invert the ${\cal F}_k''$ matrix as the number of nonzero components in $\Phi$ increases. Fortunately, the actual background field is not of particular importance when searching for the $\beta$ functions, as at each order in the field expansion one is free to choose the background at one's disposal, under the condition that the operators generated at the order in question are distinguishable. This lets the corresponding $\beta$ functions be acquired uniquely.

A potential sequence of background field choices is the following. At ${\cal O}(\phi^2)$ and ${\cal O}(\phi^3)$ we choose $\Phi=\sigma_0T_0$, which uniquely determines $\beta_{m^2}$ and $\beta_a$. At ${\cal O}(\phi^4)$ the latter background yields $\beta_{g_1}$, but since $I_2|_{\Phi=\sigma_0T_0}=0$, it is not suitable for obtaining $\beta_{g_2}$. Note that once $\beta_{g_1}$ is calculated, we may switch to $\Phi=\sigma_8 T_8$ and by subtracting the contributions of $\beta_{g_1}$ we arrive at $\beta_{g_2}$. At ${\cal O}(\phi^5)$ we switch back to $\Phi=\sigma_0T_0$ to get $\beta_{b}$. The most complicated part of the calculation occurs at ${\cal O}(\phi^6)$, as there are four operators contributing [see (\ref{Eq:F})], and we need a sequence of background field choices that uniquely allows for disentangling them. If we start with $\Phi = i\pi_0T_0$, then it yields $\beta_{\lambda_1}$ as the other three operators vanish in this background. Using the result for $\beta_{\lambda_1}$, we may switch back once again to $\Phi=\sigma_0 T_0$, which albeit mixes $\beta_{\lambda_1}$ and $\beta_{a_2}$ (but eliminates the other two invariants), since the former is already known and can be subtracted,  the latter is obtained uniquely. As for the calculation of $\beta_{\lambda_2}$ and $\beta_{g_3}$, we always need a two-component background,  as all choices for one-component fields lead the remaining two invariants to either vanish simultaneously, or they both acquire nonzero values, making them and the corresponding $\beta$ functions  impossible to disentangle. A convenient choice could be $\Phi=\sigma_8T_8 + i \pi_0 T_0$, which readily allows the unique distinction between $\beta_{\lambda_2}$ and $\beta_{g_3}$.

As opposed to the $\beta$ functions themselves, the outlined set of choices for the background fields is not unique. In principle if ${\cal F}_k''$ was invertible for a generic background field of $\Phi$, the whole approach would be to just simply expand the rhs of (\ref{Eq:flow}) in terms of $\phi^a$ and let all invariants naturally be built up while reading off their prefactors as the $\beta$ functions. Since from a practical point of view this is not possible, the outlined procedure is one of the simplest ways to get the $k$ evolution of ${\cal F}$. All $\beta$ functions are listed in the Appendix.

Similarly to the procedure of \cite{papenbrock95}, first we search for the zeros of the $\beta$ functions of the marginal interactions, i.e.,  we solve the $\beta_{\lambda_1}=0$, $\beta_{\lambda_2}=0$, $\beta_{a_2}=0$, $\beta_{g_3}=0$ equations for $\bar{\lambda}_{1,k}$, $\bar{\lambda}_{2,k}$, $\bar{a}_{2,k}$ and $\bar{g}_{3,k}$ in terms of the relevant couplings, $\bar{m}^2_k$, $\bar{a}_k$, $\bar{g}_{1,k}$, $\bar{g}_{2,k}$, $\bar{b}_k$. Then, after plugging the former into the $\beta$ functions of the latter couplings, we get expressions in terms of $\bar{g}_{1,k}$ and $\bar{g}_{2,k}$ that are genuinely nonperturbative. This shows that via a perturbative RG around $d=4$ no such results can be obtained.  Finally, we are ready to analyze numerically the RG flows in the five-dimensional space of $\{\bar{m}_k^2,\bar{a}_k,\bar{g}_{1,k},\bar{g}_{2,k},\bar{b}_k\}$. Note that the fixed-point equations are symmetric under the simultaneous reflections of $a \rightarrow -a$, and $b \rightarrow -b$; therefore, without the loss of generality, we assume that $a<0$.

\begin{table}[t]
\vspace{0.2cm}
  \begin{tabular}{ c | c | c | c | c | c }
    $\bar{m}^2$ & $\bar{g}_1$ & $\bar{g}_2$ & $\bar{a}$ & $\bar{b}$  & \# of RD \\ \hline
    0 & 0 & 0 & 0 & 0 & 5 \\ \hline
    -0.31496 & 0.43763 & 0 & 0 & 0 & 3 \\ \hline
    -0.38262 & 0.59726 & -0.62042 & 0 & 0 & 2 \\ \hline
    -0.01786 & 0.09163 & -0.14148 & -0.11900 & 0.39087 & 4 \\ \hline
  \end{tabular}
  \caption{Fixed points having a stability matrix with real eigenvalues. In the last column we indicate the number of relevant directions (RD).}
\end{table}

\begin{figure}[t]
\includegraphics[bb = 0 40 665 550,scale=0.3,angle=0]{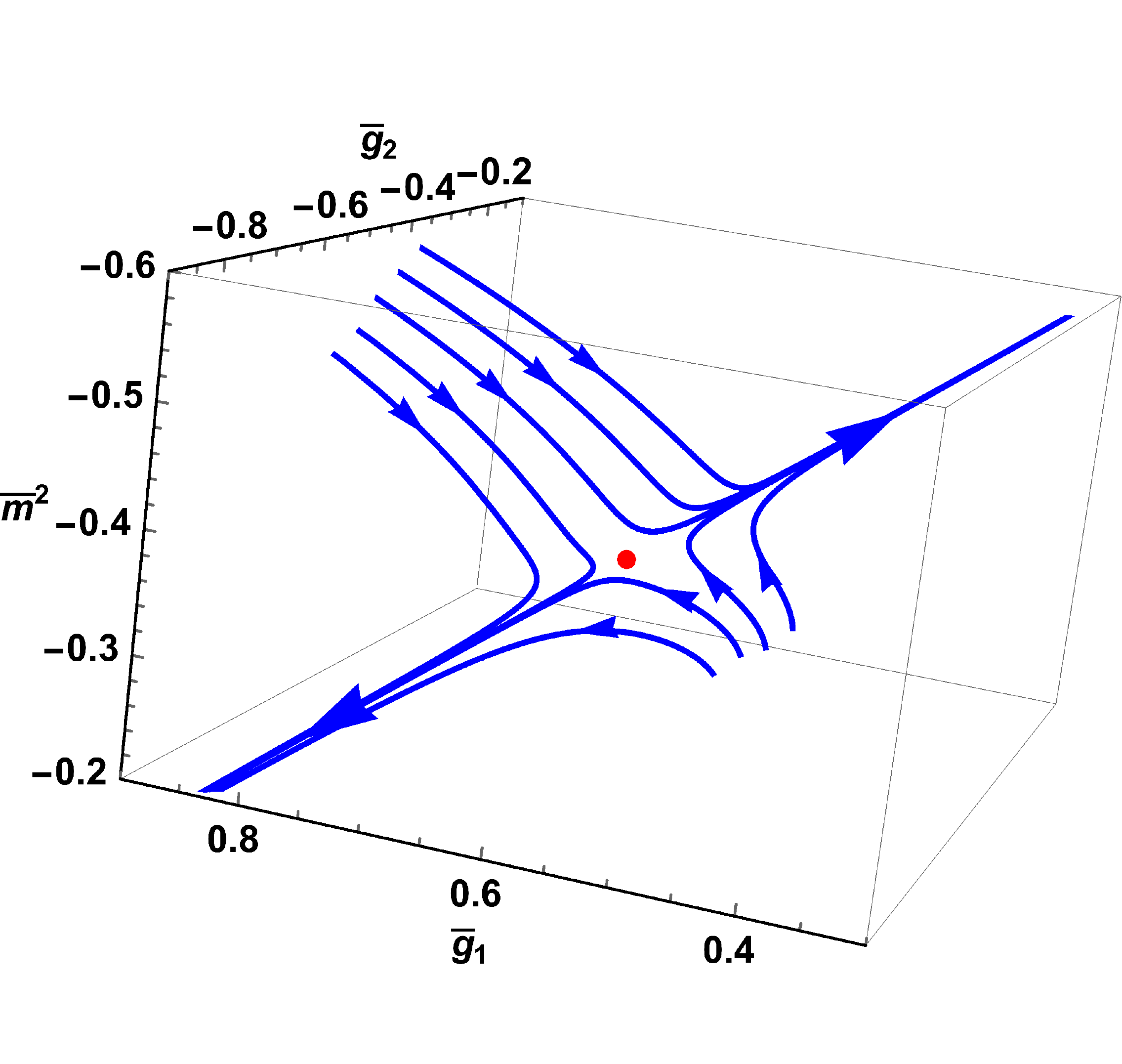}
\caption{Infrared fixed point having one relevant direction without the axial anomaly.}
\label{Fig:FP}
\end{figure}  

In Table I, we show those fixed points,\footnote{Since we are focusing on the possibility of a finite-temperature second-order transition, complex fixed points or real fixed points with complex stability eigenvalues are not listed.} which have a stability matrix, defined as $\Omega_{ij}=\partial \beta_{\omega_i} / \partial \omega_j$ (here $\omega_i$ is the collection of all couplings), with real eigenvalues. The first line is the Gaussian fixed point, the second one is the usual $O(18)$ fixed point, and in addition we find two new ones with two and four relevant directions, respectively. The one with two relevant directions is of particular interest. Since a finite temperature second-order transition corresponds to a fixed point with one relevant direction,  it seems that none of the new fixed points describes a continuous phase transition. For the one with two relevant directions, however, the stability matrix happens to be block diagonal; we have a $3\times 3$ block in the space of $\{\bar{m}^2_k,\bar{g}_{1,k},\bar{g}_{2,k}\}$,  and a $2\times 2$ block in the $\{\bar{a}_k,\bar{b}_k\}$ plane. That is, if the axial $U(1)$ symmetry is recovered at the transition point, i.e., no $\bar{a}_k$ and $\bar{b}_k$ directions exist in the Ginzburg-Landau potential, then this fixed point has only one relevant direction (see Fig. 1). This is in sharp contrast with the results of the $\epsilon$ expansion, and provided that the axial anomaly disappears at the critical temperature, the newly found fixed point may indicate a second-order chiral transition. On the flip side, one can also argue that if the chiral transition is shown to be indeed of second-order \cite{cuteri21,dini22}, then our result may hint that the axial $U(1)$ symmetry is recovered at the transition point. Note that this is the exact opposite as the two-flavor case in the $\epsilon$ expansion, where the only way to end up in a fixed point [i.e., the $O(4)$ fixed point] toward the IR is to have a large (strictly speaking infinity) initial anomaly coefficient \cite{grahl13}.

The second-order transition described by the anomaly-free IR fixed point cannot belong to any of the $O(N)$ universality classes, since the symmetry of the free energy in the aforementioned fixed point is $U(3)\times U(3)$. As for the corresponding critical exponents, the eigenvalue of the stability matrix that corresponds to the temperature variable is $y_t \approx -1.206$, which predicts the $\nu$ exponent of the transition to be $\nu =-1/y_t \approx 0.829$. Since in the present approximation no wave-function renormalization is taken into account, the $\eta$ exponent (i.e., the anomalous dimension) is zero.

\section{Conclusions}

In this paper we have calculated the renormalization group flows of all couplings up to sixth order in the three-dimensional Ginzburg-Landau potential for the three-flavor chiral transition in the zero quark mass limit. On the contrary to the results of the $\epsilon$ expansion \cite{pisarski84,butti03} and several studies using FRG flows \cite{fukushima11,grahl13,resch19,fejos14}, we find a fixed point in the infrared, which could potentially correspond to a continuous chiral transition. We believe that the discrepancy between the results of the present study and that of the aforementioned earlier works is that here we significantly extend the space of operators included in the free-energy functional, guided by the principle of (perturbative) renormalizability. As a result, in the multidimensional space of coupling constants new fixed points can be revealed, which are inaccessible in simpler truncations.

We have also found that stability requires the anomalously broken $U(1)$ axial symmetry to restore at the critical temperature; otherwise, the transition would presumably be of first-order. Our results may resolve the conflict between renormalization group arguments and recent lattice simulations \cite{cuteri21,dini22}, which predict the transition to be of second-order. If the transition is indeed continuous, our results also hint that the axial $U(1)$ symmetry is recovered at the transition point.

It would be important to investigate the robustness of the obtained results with respect to improving the truncation of the free-energy functional. There are at least three directions that are worth more exploration: (1) the inclusion of higher-order (nonrenormalizable) operators in terms of the chiral field, (2) the introduction of a field-dependent wave-function renormalization factor, and (3) taking into account higher derivative terms. These directions are already under investigation and will be reported elsewhere.

\section*{Acknowledgments} 

The author is grateful the Yukawa Institute for Theoretical Physics at Kyoto University, where this work was initiated during the YITP-W-21-09 ,,QCD phase diagram and lattice QCD" workshop. This research was supported by the Hungarian National Research, Development and Innovation Fund under Project No. PD127982, the János Bolyai Research Scholarship of the Hungarian Academy of Sciences, and the ÚNKP-21-5 New National Excellence Program of the Ministry for Innovation and Technology from the source of the National Research, Development and Innovation Fund.

\newpage
\begin{widetext}
\section*{Appendix: List of $\beta$ functions}

Here we list all the $\beta$ functions that were used to calculate the location and the stability of the fixed points in the system. Note that all couplings are dimensionless, rescaled by appropriate powers of the scale ($k$). For (perturbatively) relevant interactions we get
\vspace{0.5cm}
\begin{subequations}
\bea
\beta_{m^2}\equiv k\partial_k \bar{m}_k^2 &=& -2\bar{m}_k^2-\frac{4}{9\pi^2}\frac{15\bar{g}_{1,k}+4\bar{g}_{2,k}}{(1+\bar{m}_k^2)^2}+\frac{4}{3\pi^2}\frac{\bar{a}_k^2}{(1+\bar{m}_k^2)^3}, \nonumber\\
\beta_{a}\equiv k\partial_k \bar{a}_k &=& -\frac{3\bar{a}_k}{2}-\frac{4}{\pi^2}\frac{\bar{b}_k}{(1+\bar{m}_k^2)^2}+\frac{4}{3\pi^2}\frac{\bar{a}_k(3\bar{g}_{1,k}-4\bar{g}_{2,k})}{(1+\bar{m}_k^2)^3}, \nonumber\\
\beta_{g_1}\equiv k\partial_k \bar{g}_{1,k} &=& -\bar{g}_{1,k}-\frac{1}{9\pi^2}\frac{2\bar{a}_{2,k}+99\bar{\lambda}_{1,k}+16\bar{\lambda}_{2,k}}{(1+\bar{m}_k^2)^2}+\frac{4}{27\pi^2}\frac{24\bar{a}_k\bar{b}_k+117\bar{g}_{1,k}^2+48\bar{g}_{1,k}\bar{g}_{2,k}+16\bar{g}_{2,k}^2}{(1+\bar{m}_k^2)^3}\nonumber\\
&-&\frac{16}{9\pi^2}\frac{\bar{a}_k^2(6\bar{g}_{1,k}+\bar{g}_{2,k})}{(1+\bar{m}_k^2)^4}+\frac{8}{9\pi^2}\frac{\bar{a}_k^4}{(1+\bar{m}_k^2)^5},\nonumber\\
\beta_{g_2}\equiv k\partial_k \bar{g}_{2,k} &=& -\bar{g}_{2,k}+\frac{1}{3\pi^2}\frac{\bar{a}_{2,k}-5\bar{g}_{3,k}-13\bar{\lambda}_{2,k}}{(1+\bar{m}_k^2)^2}-\frac{4}{3\pi^2}\frac{\bar{a}_k\bar{b}_k-6\bar{g}_{1,k}\bar{g}_{2,k}-4\bar{g}_{2,k}^2}{(1+\bar{m}_k^2)^3}+\frac{4}{3\pi^2}\frac{\bar{a}_k^2(3\bar{g}_{1,k}+5\bar{g}_{2,k})}{(1+\bar{m}_k^2)^4}\nonumber\\
&+&\frac{2}{3\pi^2}\frac{\bar{a}_k^4}{(1+\bar{m}_k^2)^5}, \nonumber\\
\beta_{b} \equiv k\partial_k \bar{b}_k &=& -\frac{\bar{b}_k}{2} + \frac{4}{9\pi^2}\frac{\bar{b}_k(66\bar{g}_{1,k}-4\bar{g}_{2,k})+3\bar{a}_k(5\bar{a}_{2,k}+9\bar{\lambda}_{1,k}-4\bar{\lambda}_{2,k})}{(1+\bar{m}_k^2)^3}\nonumber\\
&+&\frac{8}{3\pi^2}\frac{-3\bar{a}_k^2\bar{b}_k-18\bar{a}_k\bar{g}_{1,k}^2+12\bar{a}_k\bar{g}_{1,k}\bar{g}_{2,k}+4\bar{a}_k\bar{g}_{2,k}^2}{(1+\bar{m}_k^2)^4}+\frac{32}{9\pi^2}\frac{\bar{a}_k^3(3\bar{g}_{1,k}-\bar{g}_{2,k})}{(1+\bar{m}_k^2)^5},  \nonumber
\eea
\end{subequations}
while for the marginal ones we obtain\vspace{-0.02cm}
\begin{subequations}
\bea
\beta_{\lambda_{1}}&\equiv& k\partial_k \bar{\lambda}_{1,k} = \frac{8}{27\pi^2}\frac{9\bar{b}_k^2+3\bar{a}_{2,k}\bar{g}_{1,k}+24\bar{g}_{1,k}(9\bar{\lambda}_{1,k}+\bar{\lambda}_{2,k})+4\bar{g}_{2,k}(9\bar{\lambda}_{1,k}+4\bar{\lambda}_{2,k})}{(1+\bar{m}_k^2)^3}\nonumber\\
&-&\frac{4}{81\pi^2}\frac{72\bar{a}_k\bar{b}_k(9\bar{g}_{1,k}+\bar{g}_{2,k})+4(297\bar{g}_{1,k}^3+108\bar{g}_{1,k}^2\bar{g}_{2,k}+72\bar{g}_{1,k}\bar{g}_{2,k}^2+16\bar{g}_{2,k}^3)+9\bar{a}^2_k(2\bar{a}_{2,k}+45\bar{\lambda}_{1,k}+4\bar{\lambda}_{2,k})}{(1+\bar{m}_k^2)^4}\nonumber\\
&+&\frac{32}{81\pi^2}\frac{\bar{a}_k^2(15\bar{a}_k\bar{b}_k+171\bar{g}_{1,k}^2+36\bar{g}_{1,k}\bar{g}_{2,k}+8\bar{g}_{2,k})}{(1+\bar{m}_k^2)^5}-\frac{80}{81\pi^2}\frac{\bar{a}_k^4(15\bar{g}_{1,k}+\bar{g}_{2,k})}{(1+\bar{m}_k^2)^6}+\frac{8}{9\pi^2}\frac{\bar{a}_k^6}{(1+\bar{m}_k^2)^7}, \nonumber\\
\beta_{\lambda_{2}}&\equiv& k\partial_k \bar{\lambda}_{2,k}=\frac{2}{9\pi^2}\frac{2\bar{g}_{2,k}(25\bar{g}_{3,k}+54\bar{\lambda}_{1,k}+44\bar{\lambda}_{2,k}-2\bar{a}_{2,k})-9\bar{b}_k^2-6\bar{g}_{1,k}(\bar{a}_{2,k}-5\bar{g}_{3,k}-28\bar{\lambda}_{2,k})}{(1+\bar{m}_k^2)^3}\nonumber\\
&+&\frac{1}{3\pi^2}\frac{36\bar{a}_k\bar{b}_k(2\bar{g}_{1,k}+\bar{g}_{2,k})-8\bar{g}_{2,k}(36\bar{g}_{1,k}^2+21\bar{g}_{1,k}\bar{g}_{2,k}+7\bar{g}_{2,k}^2)+\bar{a}_k^2(6\bar{a}_{2,k}+5\bar{g}_{3,k}+36\bar{\lambda}_{1,k})}{(1+\bar{m}_k^2)^4}\nonumber\\
&+&\frac{8}{27\pi^2}\frac{9\bar{a}_k^3\bar{b}_k+180\bar{a}_k^2\bar{g}^2_{1,k}+132\bar{a}_k^2\bar{g}_{1,k}\bar{g}_{2,k}+26\bar{a}_k^2\bar{g}^2_{2,k}}{(1+\bar{m}_k^2)^5}+\frac{20}{9\pi^2}\frac{\bar{a}_k^4(3\bar{g}_{1,k}+2\bar{g}_{2,k})}{(1+\bar{m}_k^2)^6},\nonumber\\
\beta_{a_2}&\equiv& k\partial_k \bar{a}_{2,k} = \frac{4}{3\pi^2}\frac{6\bar{b}_k^2+15\bar{a}_{2,k}\bar{g}_{1,k}-8\bar{a}_{2,k}\bar{g}_{2,k}}{(1+\bar{m}_k^2)^3}+\frac{16}{\pi^2}\frac{\bar{a}_k\bar{b}_k(\bar{g}_{2,k}-3\bar{g}_{1,k})}{(1+\bar{m}_k^2)^4}+\frac{16}{3\pi^2}\frac{\bar{a}_k^2(9\bar{g}_{1,k}^2+2\bar{g}_{2,k})}{(1+\bar{m}_k^2)^5}, \nonumber\\
\beta_{g_3}&\equiv & k\partial_k \bar{g}_{3,k} = \frac{4}{3\pi^2}\frac{15\bar{g}_{1,k}\bar{g}_{3,k}+\bar{g}_{2,k}(2\bar{a}_{2,k}+\bar{g}_{3,k}+12\bar{\lambda}_{2,k})}{(1+\bar{m}_k^2)^3}\nonumber\\
&+&\frac{1}{\pi^2}\frac{4\bar{a}_k\bar{b}_k\bar{g}_{2,k}+8\bar{g}_{2,k}^2(\bar{g}_{2,k}-9\bar{g}_{1,k})+\bar{a}_k^2(\bar{g}_{3,k}+8\bar{\lambda}_{2,k}-2\bar{a}_{2,k})}{(1+\bar{m}_k^2)^4}+\frac{16}{9\pi^2}\frac{3\bar{a}^3_k\bar{b}_k+2\bar{a}_k^2\bar{g}_{2,k}(7\bar{g}_{2,k}-12\bar{g}_{1,k})}{(1+\bar{m}_k^2)^5}\nonumber\\
&+&\frac{20}{9\pi^2}\frac{\bar{a}_k^4(5\bar{g}_{2,k}-6\bar{g}_{1,k})}{(1+\bar{m}_k^2)^6}+\frac{2}{\pi^2}\frac{\bar{a}_k^6}{(1+\bar{m}_k^2)^7}.\nonumber
\eea
\end{subequations}
\vspace{-1.15cm}
\end{widetext}

\end{document}